\documentclass[twocolumn]{aastex62}
\pdfoutput=1 
\usepackage{amsmath,amstext}
\usepackage[T1]{fontenc}
\usepackage{apjfonts} 
\usepackage{multirow}
\usepackage{braket}
\usepackage[figure,figure*]{hypcap}

\newcommand{\avir}{\mbox{$\alpha_{\rm vir}$}}
\newcommand{\aco}{\mbox{$\alpha_{\rm CO}$}}
\newcommand{\sigmol}{\mbox{$\Sigma_{\rm mol}$}}

\newcommand{\eff}{\mbox{$\epsilon_{\rm ff}$}}
\newcommand{\tff}{\mbox{$\tau_{\rm ff}$}}
\newcommand{\tdep}{\mbox{$\tau_{\rm dep}^{\rm mol}$}}
\newcommand{\acounits}{\mbox{M$_\odot$ pc$^{-2}$ (K km s$^{-1}$)$^{-1}$}}
\newcommand{\tdepkpc}{\mbox{$\tau_{\rm dep,\,1.3kpc}^{\rm mol}$}}
\newcommand{\sigmolkpc}{\mbox{$\Sigma_{\rm mol, 1.3 kpc}$}}
\newcommand{\sigsfrkpc}{\mbox{$\Sigma_{\rm SFR, 1.3 kpc}$}}
\newcommand{\sigmolpc}{\mbox{$\Sigma_{\rm mol, 60 pc}$}}

\newcommand{\tffavg}{\mbox{$\left<\tff_{\rm ,60pc}\right>_{\rm 1.3 kpc}$}}
\newcommand{\effavg}{\mbox{$\left<\epsilon_{\rm ff,60pc}\right>_{\rm 1.3 kpc}$}}

\shorttitle{Star Formation Efficiency per Free Fall Time in Nearby Galaxies}
\shortauthors{Utomo et al.}

\begin{document}

\title{THE STAR FORMATION EFFICIENCY PER FREE FALL TIME IN NEARBY GALAXIES}

\author{Dyas Utomo}
\author{Jiayi Sun}
\author{Adam K. Leroy}
\affiliation{Department of Astronomy, The Ohio State University, 140 West 18th Ave, Columbus, OH 43210, USA; utomo.6@osu.edu}
\author{J. M. Diederik Kruijssen}
\affiliation{Astronomisches Rechen-Institut, Zentrum f{\"u}r Astronomie der Universit{\"a}t Heidelberg, M{\"o}nchhofstra{\ss}e 12-14, D-69120 Heidelberg, Germany}
\author{Eva Schinnerer}
\affiliation{Max-Planck-Institut f{\"u}r Astronomie, K{\"o}nigstuhl 17, D-69117 Heidelberg, Germany}
\author{Andreas Schruba}
\affiliation{Max-Planck-Institut f{\"u}r Extraterrestrische Physik, Giessenbachstra{\ss}e 1, D-85748 Garching bei Muenchen, Germany}
\author{Frank Bigiel}
\affiliation{Institut f{\"u}r Theoretische Astrophysik, Zentrum f{\"u}r Astronomie der Universit{\"a}t Heidelberg, Albert-Ueberle-Stra{\ss}e 2, 69120 Heidelberg, Germany}
\author{Guillermo A. Blanc}
\affiliation{The Observatories of the Carnegie Institution for Science, 813 Santa Barbara Street, Pasadena, CA 91101, USA}
\affiliation{Departamento de Astronom{\'i}a, Universidad de Chile, Casilla 36-D, Santiago, Chile}
\author{M{\'e}lanie Chevance}
\affiliation{Astronomisches Rechen-Institut, Zentrum f{\"u}r Astronomie der Universit{\"a}t Heidelberg, M{\"o}nchhofstra{\ss}e 12-14, D-69120 Heidelberg, Germany}
\author{Eric Emsellem}
\affiliation{European Southern Observatory, Karl-Schwarzschild Stra{\ss}e 2, D-85748 Garching bei M{\"u}nchen, Germany}
\author{Cinthya Herrera}
\affiliation{IRAM, 300 rue de la Piscine, F-38406 Saint Martin d'H{\'e}res, France}
\author{Alexander P. S. Hygate}
\affiliation{Max-Planck-Institut f{\"u}r Astronomie, K{\"o}nigstuhl 17, D-69117 Heidelberg, Germany}
\affiliation{Astronomisches Rechen-Institut, Zentrum f{\"u}r Astronomie der Universit{\"a}t Heidelberg, M{\"o}nchhofstra{\ss}e 12-14, D-69120 Heidelberg, Germany}
\author{Kathryn Kreckel}
\affiliation{Max-Planck-Institut f{\"u}r Astronomie, K{\"o}nigstuhl 17, D-69117 Heidelberg, Germany}
\author{Eve C. Ostriker}
\affiliation{Department of Astrophysical Sciences, Princeton University, Peyton Hall, 4th Ivy Lane, Princeton, NJ 08544, USA}
\author{Jerome Pety}
\affiliation{IRAM, 300 rue de la Piscine, F-38406 Saint Martin d'H{\'e}res, France}
\affiliation{Sorbonne Universit{\'e}, Observatoire de Paris, Universit{\'e} PSL, {\'E}cole normale sup{\'e}rieure, CNRS, LERMA, F-75005, Paris, France}
\author{Miguel Querejeta}
\affiliation{European Southern Observatory, Karl-Schwarzschild Stra{\ss}e 2, D-85748 Garching bei M{\"u}nchen, Germany}
\affiliation{Observatorio Astron{\'o}mico Nacional (OAN), C/Alfonso XII 3, Madrid E-28014, Spain}
\author{Erik Rosolowsky}
\affiliation{Department of Physics, University of Alberta, 4-183 CCIS, Edmonton, AB T6G 2E1, Canada}
\author{Karin M. Sandstrom}
\affiliation{Center for Astrophysics and Space Sciences, Department of Physics, University of California, San Diego, 9500 Gilman Drive, La Jolla, CA 92093, USA}
\author{Antonio Usero}
\affiliation{Observatorio Astron{\'o}mico Nacional (OAN), C/Alfonso XII 3, Madrid E-28014, Spain}

\begin{abstract}
We estimate the star formation efficiency per gravitational free fall time, \eff, from observations of nearby galaxies with resolution matched to the typical size of a Giant Molecular Cloud. This quantity, \eff, is theoretically important but so far has only been measured for Milky Way clouds or inferred indirectly in a few other galaxies. Using new, high resolution CO imaging from the PHANGS-ALMA survey, we estimate the gravitational free-fall time at 60 to 120 pc resolution, and contrast this with the local molecular gas depletion time to estimate \eff. Assuming a constant thickness of the molecular gas layer ($H = 100$ pc) across the whole sample, the median value of \eff\ in our sample is $0.7\%$. We find a mild scale-dependence, with higher \eff\ measured at coarser resolution. Individual galaxies show different values of \eff , with the median \eff\ ranging from $0.3\%$~to~$2.6\%$. We find the highest \eff\ in our lowest mass targets, reflecting both long free-fall times and short depletion times, though we caution that both measurements are subject to biases in low mass galaxies. We estimate the key systematic uncertainties, and show the dominant uncertainty to be the estimated line-of-sight depth through the molecular gas layer and the choice of star formation tracers.
\end{abstract}


\section{Introduction}

Star formation is ``inefficient,'' meaning that the star formation rate is low compared to what would be expected if cold gas collapsed directly into stars \citep[see review by][]{MCKEE07, KRUMHOLZ14}. Theoretical models of star formation in molecular clouds that attempt to explain this inefficiency include turbulent support \citep{KRUMHOLZ05,PADOAN12}, destructive feedback \citep{MURRAY10}, magnetic fields \citep{FEDERRATH15}, and dynamical stabilization \citep{OSTRIKER10,MEIDT18}.

Over the last decade, many of these models have expressed their predictions in terms of the {\em efficiency of star formation per free fall time,} \eff. This \eff\ is the fraction of gas converted into stars per gravitational free fall time, \tff. As such, \eff\ expresses the inefficiency of star formation relative to free-fall collapse. Theoretical predictions for \eff\ on cloud scales span the range from $\sim 0.1\%$ to few times $10\%$ \citep{MCKEE07,KRUMHOLZ12,FEDERRATH12,PADOAN12,RASKUTTI16}, with higher values possible for clouds with active star formation \citep{MURRAY11,LEE16} or the densest parts of clouds \citep{EVANS14}.  From numerical simulations, \eff\ increases strongly from low values in unbound gas to high values when the virial parameter is near unity \citep{PADOAN12}.  

In spite of the fact that \eff\ is the central prediction of many current models of star formation, observational constraints on this quantity have remained challenging. The issue is that \tff\ depends on the volume density of the gas, $\rho$, via

\begin{equation} 
\label{eq:tff}
\tff = \sqrt{\frac{3\pi}{32 G \rho }}~,
\end{equation}
and it is difficult to directly measure $\rho$ at cloud scales. This requires either high resolution imaging or density-sensitive multi-line spectroscopy \citep{GAO04,LEROY17B}.

Indirect estimates of \eff\ are common. For example, \citet{MURRAY11}, \citet{EVANS14}, \citet{LEE16}, and \citet{VUTI16} estimated \eff~$\approx 0.4-1.6\%$ for populations of star-forming clouds in the Milky Way (MW), and \citet{BARNES17} obtained \eff\ $\approx 1-4\%$ in the central few hundred parsec of the MW. \citet{OCHSENDORF17} extended such studies to the Large Magellanic Cloud, where they found \eff\ in the range of $12-25\%$ (depending on the adopted SFR tracer) and showed that \eff\ decreases with increasing cloud mass. The above findings for \eff\ are mean values; all of the above studies of individual GMCs \citep[as well as earlier work by][]{MOONEY88} showed a large range of efficiency, much of which maybe due to cloud's evolution. \citet{LEROY17} estimated \eff\ $\approx 0.30-0.36\%$ in M51, based on the PAWS survey \citep{SCHINNER13}, and Schruba et al. (2018b) found \eff~$\approx 0.1-1\%$ in the MW and 7 nearby galaxies. However, we still lack a statistically significant sample of \eff\ across the local galaxy population.

The most general measurement to date comes from observations of dense gas, as traced by high critical density line emission \citep[e.g., HCN;][]{GAO04B}. By equating the mean gas density of an emission line with its critical density, and adopting a dense gas conversion factor, they can infer \eff. This approach has been taken by \citet{KRUMHOLZ07} and \citet{GARCIABURILLO12}, who concluded that \eff\ is approximately constant ($0.5-1\%$). Subsequently, numerous other studies \citep{LONGMORE13,KRUIJSSEN14B,USERO15,BIGIEL16,GALLAGHER18} have used similar techniques to find an environmentally-dependent \eff\ ($0.2-4\%$).

The PHANGS\footnote{\url{http://www.phangs.org}} collaboration is now using ALMA to map the molecular gas in 74 nearby galaxies with resolution matched to the scale of an individual Giant Molecular Cloud. These observations recover the surface density of molecular gas at high physical resolution, which is closely related to the mean volume density. In this Letter, we combine the first $11$ CO(2--1) maps from PHANGS-ALMA with three CO maps from the literature. From these maps, we infer $\tff$ and compare it to the measured gas depletion time to estimate \eff . This yields the largest and most direct sample of extragalactic \eff\ measurements to date. After describing our data in $\S$\ref{sec:data} and explaining our methodology in $\S$\ref{sec:methods}, we present the key results in $\S$\ref{sec:results} and summarize them in $\S$\ref{sec:summary}.

\section{Data} \label{sec:data}

\subsection{Molecular Gas} \label{sec:mol}

We estimate molecular gas surface density from PHANGS-ALMA CO~(2--1) data for $11$ targets and archival CO data for M31 \citep[A. Schruba et al. in preparation;][]{CALDUPRIMO16}, M33 \citep{DRUARD14}, and M51 \citep{SCHINNER13}. PHANGS-ALMA uses ALMA's 12m, 7m, and total power antennas to map CO~(2--1) emission from nearby ($d \lesssim 17$~Mpc) galaxies at native angular resolution of $1-1.5\arcsec$. This translates to native physical resolutions of $\sim~60{-}120$~pc depending on the distance to the target. At their native resolutions, the CO data cubes have rms noise of $\sim0.1$~K per $2.5$~km~s$^{-1}$ channel. The inclusion of the ACA 7m and total power data means that we expect these maps to be sensitive to emission at all spatial scales.

The sample selection, observing strategy, reduction, and properties of the full $74$ galaxies in PHANGS-ALMA survey is presented in A.~K.~Leroy et al. (in preparation). Here, we use the first data sets, including three literature maps, where the CO surface brightness and line-width have been calculated by \citet{SUN18}. See that paper for a detailed presentation of masking, map construction, and completeness.

We adopt a fixed CO(2--1)-to-H$_2$ conversion factor $\alpha_{\rm CO}^{2-1} = 6.2$~\acounits. This combines the commonly adopted Galactic CO(1--0) conversion factor, $\alpha_{\rm CO}^{1-0} = 4.35$ \acounits\ \citep{BOLATTO13}, including the contribution from Helium, with a typical CO(2--1)/CO(1--0) line ratio of 0.7 \citep[e.g.,][]{SAKAMOTO97,LEROY13}. Then, we convert the CO(2--1) integrated intensity, $I_{\rm CO}^{2-1}$, to $\Sigma_{\rm mol}$ via:
\begin{equation}
\label{eq:alphaco}
\Sigma_{\rm mol} {\rm \left[ M_\odot~pc^{-2} \right]} = \alpha_{\rm CO}^{2-1} \ I_{\rm CO}^{2-1}~{\rm \left[ K~km~s^{-1} \right]}~,
\end{equation}

\noindent The M31 and M51 CO maps target the CO(1--0) line. For those we use  $\alpha_{\rm CO}^{1-0} = 4.35$~\acounits\ with no line ratio term. We apply inclination corrections to all measured surface densities.

Our sample includes a few low mass (down to $4~\times~10^9~M_\odot$), low metallicity galaxies. We explore the effect of a metallicity-dependent $\alpha_{\rm CO}$ on our results for these cases. The fraction of `CO-dark` molecular gas increases with decreasing metallicity, resulting in higher \aco\ \citep{BOLATTO13}. We use metallicities compiled by \citet[][their Table 5]{PILYUGIN04}, except for M33 and M51, where we adopt metallicities from \citet{RS08} and \citet{CROXALL15}, respectively, and NGC~1672, NGC~3627, and NGC~4535, for which we adopt metallicities from K. Kreckel et al. in preparation based on new VLT-MUSE observations. All metallicites are quoted at $0.4~R_{25}$. We calculate the metallicity-dependent \aco\ following the prescription of \citet{BOLATTO13}. Beyond metallicity effects, the central regions of many galaxies shows smaller \aco\ \citep{SANDSTROM13}. Our key result in this paper is weighted by area and the center covers only a few lines-of-sight, so we defer investigation of the impact of this effect to future papers.

\subsection{Recent Star Formation}
\label{sec:sfr}

We derive the star formation rate (SFR) surface density, $\Sigma_{\rm SFR}$, from WISE infrared and GALEX UV maps (A.~K.~Leroy et al. in preparation). The WISE maps are derived from the unWISE reprocessing of \citet{LANG14}. The GALEX maps are coadded, convolved, background subtracted maps constructed from the full-mission GALEX archive \citep[][]{MARTIN05}. We correct the FUV and NUV maps for Galactic extinction using $E(B-V)$ from the map of \citet{SCHLEGEL98} converted to the GALEX bands using the $R_{\rm NUV}$ and $R_{\rm FUV}$ values from \citet{PEEK13}. Both sets of maps are convolved to have matched Gaussian beams ($15''$ FWHM, which corresponds to $1.3$~kpc at our most distant target) and background-subtracted using control regions outside the galaxy.

We convert FUV, NUV, 12$\mu$m, and $22\mu$m intensity, $I_\nu$, to an estimate of the recent SFR using
\begin{equation}
\label{eq:sfr}
\Sigma_{\rm SFR} \ [{\rm M_\odot~yr}^{-1}{\rm kpc}^{-2}] \approx K \ I_\nu \ [{\rm MJy \ sr}^{-1}],
\end{equation}
where $K=1.04 \times 10^{-1}$, $1.04 \times 10^{-1}$, $3.77 \times 10^{-3}$, and $2.97 \times 10^{-3}$ for FUV, NUV, 12\micron, and 24\micron\ bands, respectively \citep{KENNICUTT12,JARRETT13}. We use hybrid tracers by adding the SFR derived from each choice of UV and IR band, and adopt SFR(FUV+22\micron) as a benchmark. To estimate systematic uncertainties, we test the effect of using NUV instead of FUV and using 12\micron\ instead of 22\micron.

\section{Methodology}
\label{sec:methods}

We estimate \eff\ from the ratio between the gravitational free fall time of molecular gas, \tff, and the molecular gas depletion time, $\tdep$. 

\subsection{Molecular Gas Depletion Time}
\label{sec:tdep}

We calculate \tdep\ at $1.3$~kpc resolution across each target as
\begin{equation}
\label{eq:tdep}
\tdepkpc = \frac{\sigmolkpc}{\sigsfrkpc}.
\end{equation}

\noindent Here, \sigmolkpc\ is the convolved \sigmol\ at 1.3 kpc FWHM to match the resolution of \sigsfrkpc\ maps. We treat this as our working resolution to estimate \tdep.

\subsection{Molecular Gas Free Fall Time}

We estimate \tff\ following Equation~\ref{eq:tff}. This requires an estimate of the mass volume density, $\rho$. To estimate $\rho$, we combine our measured, high physical resolutions ($60{-}120$~pc) \sigmol\ with an estimate of the line-of-sight depth through the molecular gas layer, $H$, so that:
\begin{equation} 
\label{eq:rho}
\rho \approx \frac{\sigmol}{H}~.
\end{equation}
\noindent We describe how we estimate $H$ in $\S$\ref{sec:h_values}. We combine Equations~\ref{eq:tff} and \ref{eq:rho} to estimate \tff\ as
\begin{equation}
\tau_{\rm ff, 60 pc} = \sqrt{\frac{3\pi}{32G}\left(\frac{H}{\sigmolpc}\right)}.
\end{equation}
We make analogous measurements of \tff\ at $80$, $100$, and $120$~pc resolution, as permitted by the native resolution of the data.

\subsection{Thickness of the Molecular Gas Layer} \label{sec:h_values}

To translate a measured molecular gas surface density into a volume density, we must estimate the line of sight depth of the molecular gas layer, $H$. We define $H$ so that $\rho = \sigmol / H$. We explore three approaches:

\begin{enumerate}

\item \textit{Fixed $H=100$ pc.} This is roughly the diameter of a large molecular cloud and a characteristic thickness (FWHM) of the molecular gas layer in the Milky Way and other galaxies \citep{HEYER15,YIM14,PETY13}. This is our default value.

\item In \textit{hydrostatic equilibrium}, the turbulent midplane pressure of molecular gas balances the vertical weight of the molecular gas column in the potential of the disk. If we consider only gas responding to the potential well defined by stars, i.e., neglecting gas self-gravity, then

\begin{equation}
H\,\equiv\,2h\approx\,\sqrt{\frac{\sigma_{\rm mol}^2~h_*}{G\,\Sigma_*}}~,
\end{equation}

\noindent following \citet{OSTRIKER10}. Here $\sigma_{\rm mol}$ is the velocity dispersion of the molecular gas, $\Sigma_*$ is the mass surface density of stars, and $h_*$ is the stellar scale height ($\rho_* = \Sigma_* / 2 h_*)$. Here, we adopt a typical $h_* = 300$~pc, use the measured line width from \citet{SUN18}, and estimate $\Sigma_*$ from the dust-corrected {\em Spitzer} $3.6\micron$ maps produced by \citet{QUEREJETA15}\footnote{In four galaxies, we currently lack {\em Spitzer} maps and use WISE $3.4\micron$ maps instead.} assuming a mass-to-light ratio of $0.5 ~M_\odot / L_\odot$ \citep{MEIDT14}. The median of $H$ under this assumption is 122 pc.

\item We assume that each beam contains one spherical, \textit{unresolved cloud in energy equipartition.} In this case, kinetic energy balances gravitational potential energy, equivalent to setting the virial parameter $\avir\,\approx\,2$ \citep{BERTOLDI92,SUN18}. We take $\avir\,\approx\,(5\sigma_{\rm mol}^2R)/(GM_{\rm mol})$ and calculate the mass in the beam from $M_{\rm mol} = \sigmol\,A$, where $A = \pi (\theta_{\rm FWHM}/2)^2 / \ln 2$ is the physical beam area. From this, we derive the cloud diameter, $2R$, via
\begin{equation} \label{eq:virial_cloud}
H\,\equiv\,2R\,\approx\,\frac{2\,\avir\,G\,\sigmol\,A}{5\,\sigma_{\rm mol}^2}.
\end{equation}
The median of $H$ under this assumption is 116 pc.
\end{enumerate}

We calculate $H$ using each method above and compare the resulting \eff\ to estimate the systematic uncertainty associated with estimating $H$. 

\subsection{Combining Scales}

\begin{figure*}
\centering
\includegraphics[width=\textwidth]{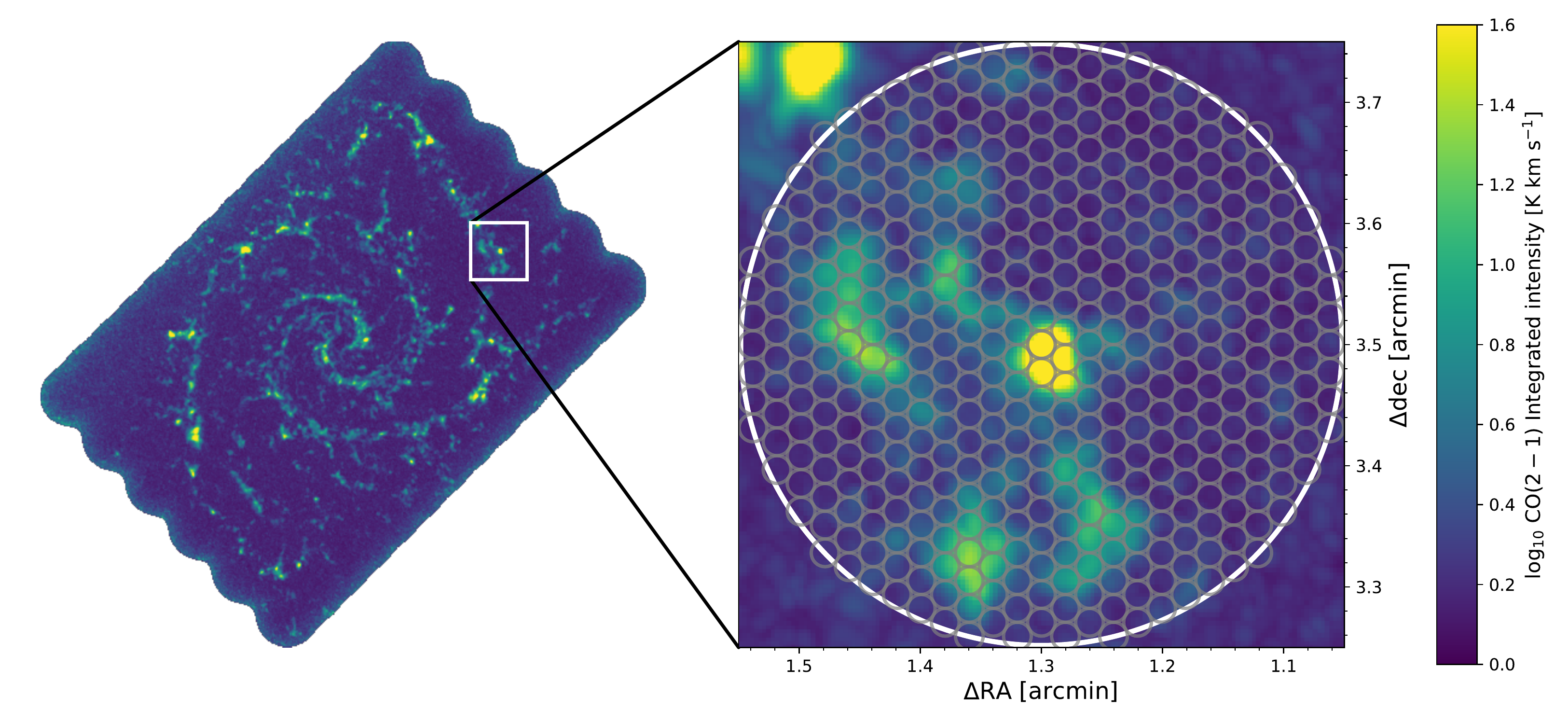}
\caption{\textit{Left:} CO(2--1) integrated intensity map of NGC~628 at $60$~pc resolution (color codes in the range of $0.0\leq{\rm log}_{10}$[CO(2--1)/K km s$^{-1}$]~$\leq1.6$). We use this map to estimate molecular gas surface density and free-fall time. \textit{Right:} Illustration of our cross-scale methodology. We measure the molecular gas depletion time, $\tau_{\rm dep} \equiv \sigmol / \Sigma_{\rm SFR}$ at $1.3$~kpc resolution (illustrated by the large circle). Within each $1.3$~kpc region, we calculate \sigmol\ and \tff\ for each 60~pc beam (small circles). We average these high resolution \tff\ estimates within 1.3~kpc region, weighted by \sigmol\ at 60~pc beam. By dividing \tff\ by \tdep, we calculate the average \eff\ within each $1.3$~kpc region while still leveraging the high resolution of the PHANGS-ALMA CO maps.}
\label{fig:map}
\end{figure*}

We estimate \tff\ at $60{-}120$~pc resolution and measure \tdep\ at 1.3 kpc resolution. To combine these measurements, we calculate the mass-weighted average of \tff\ within each $1.3$~kpc region of a galaxy. This is equivalent to asking ``What is the mass-weighted mean of \tff\ of a parcel of molecular gas in this kpc-sized region of this galaxy?'' Figure \ref{fig:map} illustrates our approach for one of our targets, NGC~628.

We calculate the mass-weighted mean of \tff$^{-1}$ via
\begin{equation}
\label{eq:avg_meas}
\left<\tau_{\rm ff, 60 pc}^{-1}\right>_{\rm 1.3 kpc} = \frac{\tau_{\rm ff, 60pc}^{-1}~\sigmolpc~\ast~\theta_{\rm 60pc}^{\rm 1.3kpc}}{\Sigma_{\rm mol, 60pc}~\ast~\theta_{\rm 60pc}^{\rm 1.3kpc}},
\end{equation}

\noindent where $\Sigma_{\rm mol, 60pc}$ is the surface density of molecular gas at $60$~pc resolution, $\theta_{\rm 60pc}^{\rm 1.3kpc}$ is the Gaussian kernel to convolve a $60$~pc resolution map to $1.3$~kpc resolution, and $\ast$ denotes convolution. We have round Gaussian beams in all maps. Hereby, we assume that $\left<\tau_{\rm ff, 60 pc}\right>_{\rm 1.3 kpc}^{-1} \approx \left<\tau_{\rm ff, 60 pc}^{-1}\right>_{\rm 1.3 kpc}$.

This differs slightly from \citet{LEROY17}. They first calculated the mass-weighted mean of {\em surface density}, and then used that to calculate \tff, instead of directly calculating the mass-weighted mean of $\tff^{-1}$. The approach here should yield a more rigorous comparison to predictions in which $\Sigma_{\rm SFR} = \eff\ \Sigma_{\rm mol}/\tff$. The two approaches yield qualitatively similar results, though, with the mean \tffavg\ differing by only $\sim 7\%$.

\subsection{Star Formation Efficiency per Free Fall Time}

We calculate \eff\ as the ratio between \tff\ and \tdep,
\begin{equation} 
\label{eq:eff}
\effavg = \frac{\left<\tau_{\rm ff, 60 pc}\right>_{\rm 1.3 kpc}}{\tdepkpc}.
\end{equation}
\noindent We carry out analogous calculations at 80, 100, and 120 pc resolutions. This allows us to study the impact of varying the linear resolution on the measured values of \eff . Our targets vary in their native physical resolutions, so not all targets are available at the highest resolutions \citep{SUN18}.

\subsection{Correction for Incompleteness}
\label{sec:caveat}

When estimating \tffavg, we begin with a high resolution map that has been masked using a signal-to-noise cut \citep{SUN18}. The calculation will miss emission at signal-to-noise below this cut, which has preferentially low $\sigmol$ and long \tff. \citet{SUN18} measured the degree of this effect for each of our maps. They define the completeness, $C$, as the fraction of the total CO flux, measured at lower resolution with very good signal-to-noise, that is included in the high resolution, masked map. For our targets, $C$ ranges from 44\% to 96\% at 120 pc resolution, and is typically lower at finer resolutions.

To estimate the effect of incompleteness on our calculated \tff, we use a Monte Carlo approach. We randomly draw $10^6$ samples from a lognormal distribution designed to simulate the true distribution of mass as a function of \sigmol\ \citep[see][]{LEROY16,SUN18}. These model distributions have $1\sigma$ width of $0.5$~dex. For each distribution, we calculate true expectation value of $\tff^{-1}$ weighted by \sigmol, for the whole distribution and for subsets of the sample where only the highest fraction $C$ of the data are included.

This yields a correction factor $f_C$, defined as the ratio of the true $\left<\tau_{\rm ff}\right>$ over the measured $\left<\tau_{\rm ff}\right>$, as a function of $C$. We apply these to the data based on the value of $C$ measured in each $1.3$~kpc larger beam \citep[our flux recovery is nearly perfect at $1.3$~kpc resolution;][]{LEROY16,SUN18}. Incompleteness suppresses faint, long $\tff$ lines-of-sight, so that $1.0 \lesssim f_C \lesssim 1.1$ for 120 pc beam. Therefore, correcting for incompleteness increases $\tff$ and \eff.

\section{Results}
\label{sec:results}

\begin{figure*}
    \centering
    \epsscale{1.18}
    \plotone{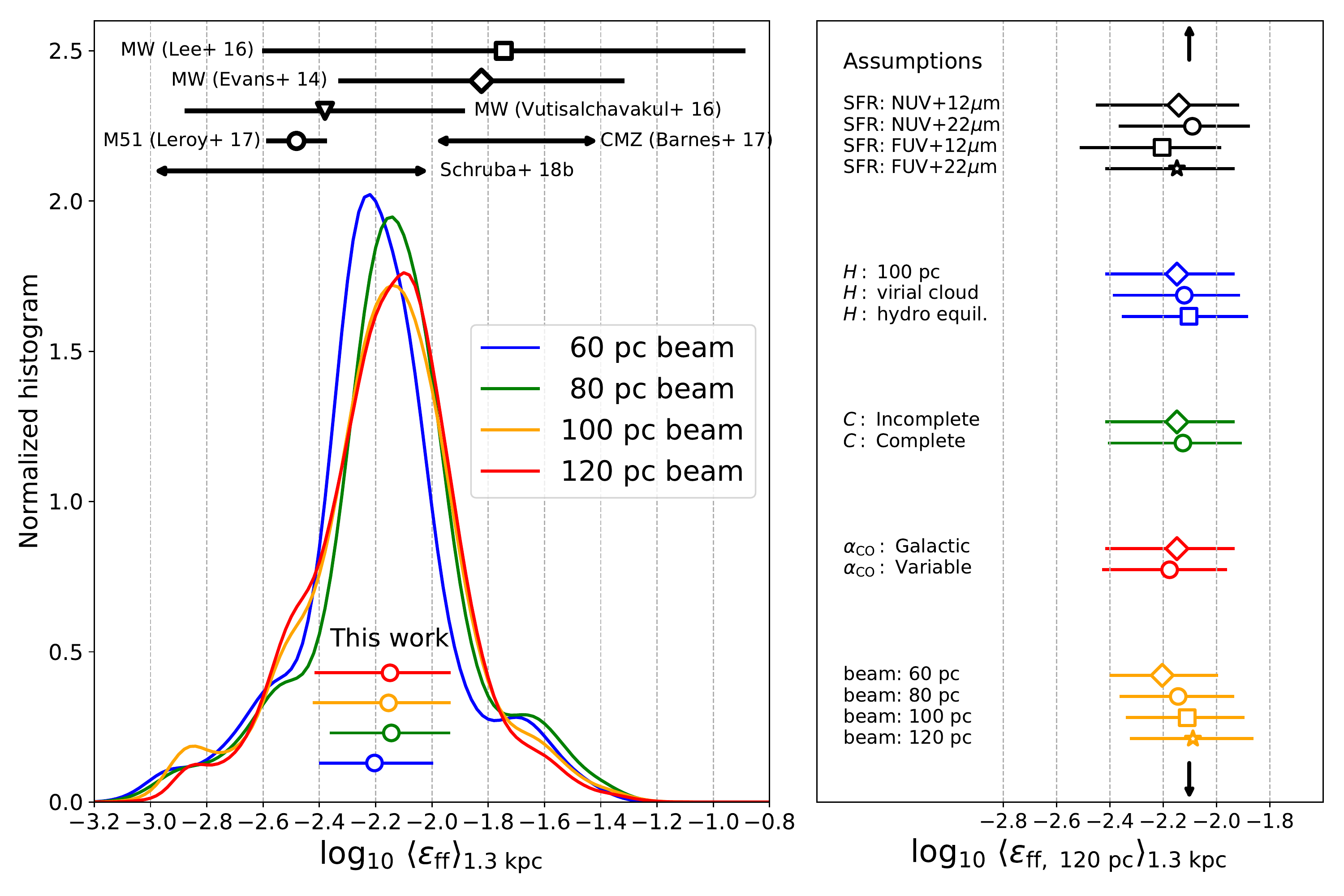}
    \caption{Efficiency per free fall time in 14 galaxies. {\it Left:} histograms of \eff\ for all regions studied (smoothed using a Gaussian kernel density estimator following \citealt{SCOTT92}). The histograms combine all galaxies and give equal weight to each $1.3$~kpc region regardless of galaxy or other properties. Different colors show results for \tff\ calculated at different resolutions. Circles and lines mark the median, 16$^{\rm th}$, and $84^{\rm th}$ percentile \eff\ for each resolution. For comparison, we show the values of \eff\ measured for Milky Way clouds \citep{EVANS14,LEE16}, in the Central Molecular Zone \citep{BARNES17}, in M51 at 40 pc resolution \citep{LEROY17}, and a compilation study for MW and 7 nearby galaxies (Schruba et al. 2018b). {\it Right:} Impact of assumptions and uncertainties. We plot the median and $16^{\rm th}-84^{\rm th}$ percentile range of $\left<\eff\right>$ by altering various assumptions from our default values at 120 pc resolution. For various linear resolutions, we only include 9 galaxies that can be resolved down to 60 pc. Most systematic uncertainties affect the results at the $\sim 0.1$~dex ($\sim 25\%)$ level. The choices of SFR tracers and beam sizes have the largest impact. We also plot the median $\left<\eff\right>$, giving equal weight for each galaxy, as small vertical arrows.}
    \label{fig:histo}
\end{figure*}

In the left panel of Figure~\ref{fig:histo} and Table~\ref{tab:stats}, we summarize our measurements of \eff\ for the whole sample, using our standard assumption ($H=100$~pc, SFR from FUV$+22\micron$, incomplete, and Galactic \aco). These measurements over a large area across $14$ galaxies represent the most complete measurement of the efficiency of star formation per free fall time to date. At $120$~pc resolution (red histogram), we find median $\eff \approx 0.7\%$ across all lines-of-sight in $14$ galaxies, with the $16{-}84\%$ percentile range spanning $\eff \approx 0.4{-}1.1\%$.

The number of lines-of-sight varies in each galaxy. If instead, we take a median value for each galaxy, and compute the overall median across the whole sample (equivalent to giving equal weight to each galaxy), then $\eff \approx 0.8\%$. Those \eff\ values are the most fundamental result of this letter.

\begin{deluxetable}{ l | c c c c }
\tablewidth{0pt}
\tabletypesize{\scriptsize}
\tablecaption{Measurements summary for all lines-of-sight. \label{tab:stats}}
\tablehead{
\\
{} & \multicolumn{4}{c}{Physical resolutions of CO maps} \\
Quantities & 60 pc & 80 pc & 100 pc & 120 pc
}
\startdata
Numbers of galaxies & 9 & 9 & 11 & 14 \\
Numbers of lines-of-sight & 949 & 949 & 1651 & 2937 \\
\hline
Median of $\left<\epsilon_{\rm ff}\right>$ & $0.63\%$ & $0.72\%$ & $0.70\%$ & $0.71\%$ \\
16$^{\rm th}$ percentile of $\left<\epsilon_{\rm ff}\right>$ & $0.40\%$ & $0.44\%$ & $0.38\%$ & $0.39\%$ \\
84$^{\rm th}$ percentile of $\left<\epsilon_{\rm ff}\right>$ & $1.00\%$ & $1.15\%$ & $1.15\%$ & $1.15\%$ \\
\hline
Median of $\left<\tau_{\rm ff}\right>$ [Myr] & 11.16 & 12.68 & 12.54 & 11.79 \\
16$^{\rm th}$ percentile of $\left<\tau_{\rm ff}\right>$ [Myr] & 6.52 & 7.04 & 7.29 & 7.57 \\
84$^{\rm th}$ percentile of $\left<\tau_{\rm ff}\right>$ [Myr] & 13.62 & 15.74 & 15.75 & 15.59 \\
\hline
Median of $\tau_{\rm dep}$ [Gyr] & 1.72 & 1.72 & 1.77 & 1.69 \\
16$^{\rm th}$ percentile of $\tau_{\rm dep}$ [Gyr] & 1.18 & 1.18 & 1.13 & 1.11 \\
84$^{\rm th}$ percentile of $\tau_{\rm dep}$ [Gyr] & 2.29 & 2.29 & 2.41 & 2.35 \\
\enddata
\end{deluxetable}

\subsection{Uncertainties} 
\label{sec:unc}

The histograms in Figure~\ref{fig:histo} combine more than $940$ regions of 1.3~kpc in size (see Table \ref{tab:stats}), and the statistical uncertainties on any given \eff\ estimate tend to be quite small ($\lesssim0.01$~dex), because many measurements are already averaged together within each $1.3$~kpc beam. As a result, we expect that the spread in the histogram to represent real physical variations in \eff\ from region to region and from galaxy to galaxy. The dominant uncertainties affecting the measurement are systematic. We explore the magnitude of these systematic uncertainties in the right panel of Figure~\ref{fig:histo}, where we vary our adopted SFR tracer, the line-of-sight depth, completeness correction, the CO-to-H$_2$ conversion factor, and linear resolution.

In general, over the range of assumptions that we explore, systematic effects can shift \eff\ by $\sim 0.1$~dex. In particular, altering our mix of SFR tracers shifts \eff\ by $\lesssim 0.1$~dex. Adopting a metallicity-dependent $\alpha_{\rm CO}$ only has a small impact on the median \eff\ of the whole sample because our low mass galaxies contribute only a small fraction of the total lines-of-sight. However, variations in $\alpha_{\rm CO}$ have a more significant impact on the measured \eff\ in individual galaxies ($\S$\ref{sec:g2g}).

Varying the resolution of the maps changes \eff, but only weakly. Within our sample, changing the resolution from 60 to 120~pc increases \eff\ by $\sim 0.1$~dex. This is consistent with the idea that beam dilution decreases the measured \sigmol\ as the resolution degrades, which in turn raises \tff\ and \eff. 
Other systematic uncertainties stem from imperfect knowledge of the disk thickness, $H$, and incompleteness due to limited sensitivity in the high resolution CO maps. The right panel of Figure \ref{fig:histo} shows that correcting for the presence of low \sigmol , high \tff\ lines-of-sight shifts \eff\ towards higher value by $< 0.1$~dex. Meanwhile, adopting different plausible treatments of $H$ can also shift \eff\ by $\lesssim 0.1$~dex. Direct measurements of the vertical distribution of the cold gas in galaxies \citep[][]{YIM11,YIM14} will help to constraint $H$ and \eff.

\subsection{Comparison to Previous Studies}

We find $\eff \approx 0.7\%\pm0.3 \%$. This value is comparable to the often-quoted theoretical values of $\approx 1\%$ \citep{MCKEE07,KRUMHOLZ07,KRUMHOLZ12}. Numerical simulations of kpc-scale regions of the ISM with star formation feedback found \eff~$\approx~0.6\%$ \citep{KIM13}; this can be understood based on expectations from UV heating and turbulence driving by supernovae \citep{OSTRIKER10,OSTRIKER11}. Our \eff\ value is  lower than $\eff \approx 10\%$ suggested by \citet{AGERTZ15}, but they also argued that their high {\it local} efficiency is derived from a short cloud-scale \tdep\ (rather than kpc-scales as in our work), and can still result in a low apparent {\it global} efficiency ($\sim0.25\%$) if a global (kpc-scales) \tdep\ of $\sim2$ Gyrs \citep[][this work]{LEROY13,UTOMO17} is adopted.

As Figure~\ref{fig:histo} shows, our measured \eff\ is low compared to the median \eff\ $\sim 1.5{-}1.8\%$ found in the Milky Way (MW) clouds by \citet{EVANS14}, \citet{LEE16}, and \citet{BARNES17}. This can be partially understood because the focus of MW measurements is on the high column density parts of clouds \citep{EVANS14} and on actively star forming clouds \citep{LEE16}. \citeauthor{EVANS14} measured \eff\ within a visual extinction contour of $A_V>2$ magnitude (equivalent to $\Sigma_{\rm mol} \gtrsim 20~M_{\odot}$~pc$^{-2}$). Our measurements also integrate over lower column density regions, resulting in \tff\ and \tdep\ $\sim8$ and 16 times longer than those in \citeauthor{EVANS14} Indeed, \citet{VUTI16} found a mean $\eff \approx 0.4\%$ by considering a sample of lower volume density of MW clouds (with mean $n_{\rm H2}\sim300$~cm$^{-3}$, instead of $800$~cm$^{-3}$ as in \citeauthor{EVANS14}).

Furthermore, we expect the difference with the \citet{LEE16} MW measurements to reflect a bias towards actively star forming clouds in their sample \citep[e.g.,][\S2.1]{KRUIJSSEN14,KRUIJSSEN18}. Their measurements include $\sim 80\%$ of the ionizing photon flux in the MW, but only captured $\sim 10\%$ of the total GMC mass in the \citet{MIVILLE-DESCHENES17} catalog. Our measurements include all CO emission in each 1.3 kpc aperture, so that clouds and star forming regions in all evolutionary states are included (as long as they are above the sensitivity limit). Following \citet{MURRAY11}, \citet{LEE16} emphasized the large scatter of \eff\ from cloud to cloud \citep[a result that goes back to][]{MOONEY88}. Our 1.3~kpc \tdep\ measurements average over many clouds and so neither contradict nor confirm their result. Cloud-by-cloud star formation rate estimates are in progress for PHANGS (e.g., K.~Kreckel et al. in preparation), and will help to test whether the observations of \citet{MURRAY11} and \citet{LEE16} indeed hold in other galaxies.

Our median $\eff \approx 0.7\%$ in the whole sample is about twice the $\eff \approx 0.30-0.36\%$ found by \citet[][]{LEROY17} using an almost identical methodology to study M51 at 40~pc resolution. M51 is also part of our sample, and our measurements for that galaxy agree well with those in \citet{LEROY17}. This appears to reflect a real difference between M51 and the rest of our sample, i.e. M51 has the lowest \eff\ of any galaxy in our sample. Following \citet{MEIDT13}, this may reflect strong gas flows in M51 that act to stabilize the gas and suppress star formation. Strong gas flows were also observed in NGC~3627 \citep{BEUTHER17}, where \eff\ is low ($\approx0.6\%$).

\begin{figure*}
    \centering
    \includegraphics[width=0.75\textwidth]{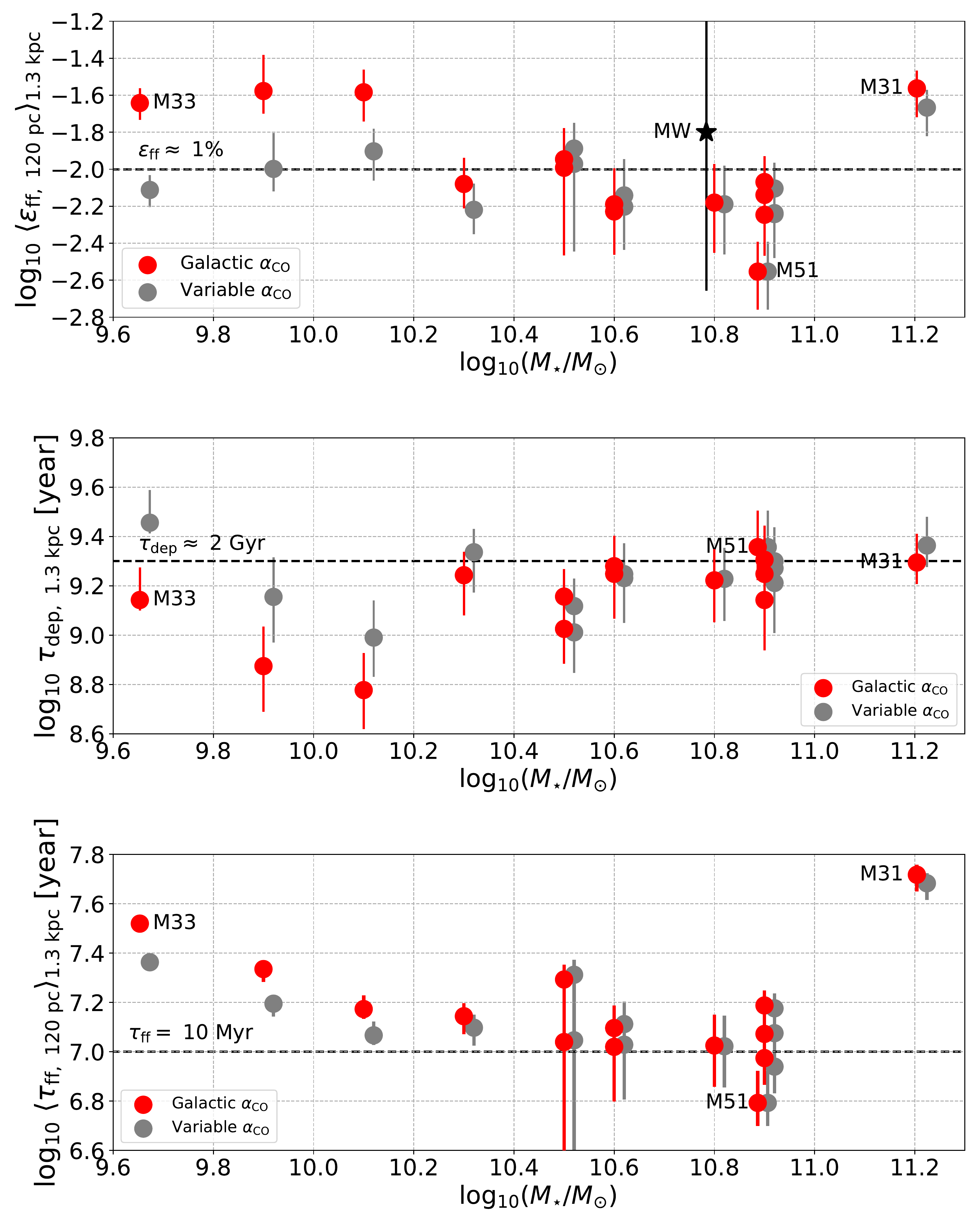}
    \caption{Galaxy-by-galaxy measurements of \eff, \tdep, and \tff. \textit{Top panel:} Median and $16^{\rm th}-84^{\rm th}$ percentile range of \eff\ for each galaxy as a function of galaxy stellar mass ($M_*$) for Galactic \aco\ (red circles and bars) and a metallicity-dependent \aco\ (gray circles and bars). We shift the gray circles to the right by 0.02~dex for clarity. \textit{Middle and bottom panels:} same as the top panel, but for \tdep\ and \tff.}
    \label{fig:gal_to_gal}
\end{figure*}

\subsection{Galaxy-to-Galaxy Variations} 
\label{sec:g2g}

Figure \ref{fig:histo} shows overall results for the whole sample, but we also observe strong galaxy-to-galaxy variations in \eff. In Figure~\ref{fig:gal_to_gal} and Table~\ref{tab:data},
we report \eff\ for each galaxy at 120~pc resolution. Red circles and bars show the median and $16^{\rm th}{-}84^{\rm th}$ percentile range for each galaxy using a Galactic \aco . Here, the contrast between low mass, low metallicity galaxies and massive galaxies stands out. To illuminate a possible cause for this, we also show results adopting metallicity-dependent \aco\ as gray circles. Because \eff\ depends on both \tdep\ and \tff, we also plot these quantities in the middle and lower panels.

The top panel of Figure~\ref{fig:gal_to_gal} shows a dynamic range of an order of magnitude in \eff\ ($\approx0.3$--$2.6\%$) across our sample. Among the high mass galaxies (excluding M31 and M51), the scatter in \eff\ is $\sim 0.2$~dex. Except for M31, \eff\ appears to decrease with increasing stellar mass of the galaxy (Spearman rank correlation coefficient, $r_s \approx -0.75$).

The middle and bottom panels show that this trend originates from a combination of changes in \tdep\ and \tff. For Galactic \aco , our three lowest mass galaxies show the shortest \tdep\ in our sample ($\lesssim 1$ Gyr). A similar \tdep$-M_\star$ trend was also observed by \citet{SAINTONGE11}, \citet{LEROY13}, and \citet{BOLATTO17}. Meanwhile, \tff\ declines with increasing stellar mass ($r_s \approx -0.64$; excluding M31). This agrees with the observation that at a fixed resolution, \sigmol\ scales with galaxy stellar mass \citep{SUN18}, leading to longer \tff\ in low mass galaxies.

Much, but not all, of the observed trends with stellar mass can be explained by the application of a metallicity dependent \aco, shown as the gray points. If a large reservoir of CO-dark molecular gas is present in these low mass galaxies \citep[e.g.,][]{LEROY11,BOLATTO13,GRATIER17,SCHRUBA17}, then \tdep\ will be longer and \tff\ shorter, resulting in lower \eff\ in the low mass galaxies. The correction that we adopt, which is uncertain, yields $\eff \sim 1\%$ in the low mass targets, similar to \eff\ in the high mass galaxies. However, even with this metallicity correction, there is still a significant anti-correlation between galaxy stellar mass and \eff\ ($r_s \approx -0.57$; excluding M31).

M31 shows a higher \eff\ that can not be explained by the metallicity-dependent \aco\ only. This apparent high efficiency may partially reflect beam-filling effects. M31 has a low molecular-to-atomic gas fraction, and if the clouds are small, widely spaced, and tenuous compared to the beam (as suggested by \citealt{SUN18}), then the long \tff\ may be partially an observational bias due to low beam filling factor.

\section{Summary} \label{sec:summary}

We estimate the star formation efficiency per gravitational free-fall time, \eff, in $14$ star-forming galaxies, where $11$ of them are part of the PHANGS-ALMA survey. This represents the most complete measurement of this key theoretical quantity across local galaxies to date. To do so, we use high resolution CO maps to infer the molecular gas volume density and free-fall time, \tff, at $60{-}120$~pc resolution. We estimate the gas depletion time from the same CO maps and archival UV and IR data, convolved to $1.3$~kpc resolution. We connect those cross-scale measurements by taking the mass-weighted average of \tff$^{-1}$ within 1.3~kpc aperture.

Overall, we find \eff\ in the range of $0.4{-}1.1\%$, with median $\approx 0.7\%$, and significant galaxy-to-galaxy scatter ($0.3-2.6\%$). We assess the impact of systematic uncertainties on this measurement to be within $0.1$~dex, with the largest uncertainties associated with the assumption of molecular gas thickness and the choice of SFR tracer. The galaxy-to-galaxy scatter in \eff\ is systematic, with an overall trend toward finding higher \eff\ in low mass galaxies and in our only ``green valley'' target, M31. We argue that these trends may be partially explained by a metallicity-dependent \aco\ and sparse, small clouds in M31.

\begin{deluxetable*}{l c c c c c c c c c c c c}
\tablewidth{0pt}
\tabletypesize{\scriptsize}
\tablecaption{Measurements for each galaxy at 120 pc beam of CO maps. \label{tab:data}}
\tablehead{
\\
Galaxies & Morphology & distance & inclination & log$_{10}M_*$ & log$_{10}$SFR & \# l.o.s & log$_{10}$\tff\ & log$_{10}$\tdep\ & log$_{10}$\eff & $C$ & $f_C$ & \aco(1--0) \\
{} & {} & Mpc & degree & $M_\odot$ & $M_\odot$ yr$^{-1}$ & {} & years & years & {} & {} & {} & (see notes)
}
\startdata
NGC~0224 & Sb-A & 0.8 & 77.7 & 11.20 & $-0.43$ & 22 & $7.72_{-0.06(-0.30)}^{+0.03(+0.18)}$ & $9.29_{-0.08(-0.14)}^{+0.11(+0.07)}$ & $-1.56_{-0.15(-0.05)}^{+0.09(+0.65)}$ & $0.97_{-0.05}^{+0.10}$ & $1.01_{-0.01}^{+0.01}$ & $5.10$ \\
NGC~6744 & Sbc-AB & 11.6 & 40.0 & 10.90 & $0.39$ & 299 & $7.19_{-0.05(-0.02)}^{+0.05(+0.04)}$ & $9.25_{-0.10(-0.06)}^{+0.11(+0.02)}$ & $-2.07_{-0.10(-0.00)}^{+0.13(+0.16)}$ & $0.67_{-0.17}^{+0.15}$ & $1.05_{-0.02}^{+0.02}$ & $4.59$ \\
NGC~4321 & Sbc-AB & 15.2 & 27.0 & 10.90 & $0.53$ & 525 & $7.07_{-0.15(-0.00)}^{+0.09(+0.04)}$ & $9.31_{-0.15(-0.06)}^{+0.13(+0.07)}$ & $-2.25_{-0.22(-0.04)}^{+0.20(+0.14)}$ & $0.65_{-0.31}^{+0.19}$ & $1.05_{-0.03}^{+0.05}$ & $4.29$ \\
NGC~4303 & Sbc-AB & 17.6 & 25.0 & 10.90 & $0.72$ & 424 & $6.97_{-0.10(-0.03)}^{+0.08(+0.08)}$ & $9.14_{-0.20(-0.06)}^{+0.14(+0.07)}$ & $-2.14_{-0.23(-0.11)}^{+0.18(+0.13)}$ & $0.75_{-0.26}^{+0.13}$ & $1.03_{-0.02}^{+0.04}$ & $5.10$ \\
NGC~5194 & Sbc-A & 7.6 & 21.0 & 10.89 & $0.43$ & 100 & $6.79_{-0.09(-0.00)}^{+0.12(+0.08)}$ & $9.36_{-0.13(-0.05)}^{+0.14(+0.10)}$ & $-2.55_{-0.20(-0.10)}^{+0.15(+0.14)}$ & $0.90_{-0.17}^{+0.06}$ & $1.01_{-0.01}^{+0.02}$ & $4.35$ \\
NGC~4254 & Sc-A & 16.8 & 27.0 & 10.80 & $0.68$ & 553 & $7.03_{-0.16(-0.00)}^{+0.12(+0.08)}$ & $9.22_{-0.17(-0.05)}^{+0.12(+0.05)}$ & $-2.18_{-0.26(-0.05)}^{+0.20(+0.13)}$ & $0.76_{-0.26}^{+0.15}$ & $1.03_{-0.02}^{+0.04}$ & $4.41$ \\
NGC~4535 & Sc-AB & 15.8 & 40.0 & 10.60 & $0.30$ & 314 & $7.10_{-0.14(-0.00)}^{+0.08(+0.07)}$ & $9.28_{-0.18(-0.06)}^{+0.12(+0.06)}$ & $-2.19_{-0.24(-0.00)}^{+0.19(+0.20)}$ & $0.63_{-0.16}^{+0.15}$ & $1.05_{-0.02}^{+0.03}$ & $4.04$ \\
NGC~3627 & Sb-AB & 8.3 & 62.0 & 10.60 & $0.10$ & 153 & $7.02_{-0.22(-0.04)}^{+0.11(+0.19)}$ & $9.25_{-0.18(-0.05)}^{+0.14(+0.13)}$ & $-2.23_{-0.23(-0.00)}^{+0.19(+0.40)}$ & $0.89_{-0.10}^{+0.06}$ & $1.02_{-0.01}^{+0.01}$ & $4.18$ \\
NGC~3351 & Sb-B & 10.0 & 41.0 & 10.50 & $0.11$ & 93 & $7.29_{-0.71(-0.08)}^{+0.05(+0.02)}$ & $9.16_{-0.27(-0.08)}^{+0.11(+0.12)}$ & $-1.95_{-0.31(-0.06)}^{+0.13(+0.24)}$ & $0.76_{-0.12}^{+0.16}$ & $1.03_{-0.02}^{+0.02}$ & $3.98$ \\
NGC~1672 & Sb-B & 11.9 & 40.0 & 10.50 & $0.48$ & 172 & $7.04_{-0.34(-0.02)}^{+0.15(+0.10)}$ & $9.03_{-0.14(-0.06)}^{+0.13(+0.13)}$ & $-1.99_{-0.47(-0.09)}^{+0.21(+0.22)}$ & $0.69_{-0.19}^{+0.17}$ & $1.04_{-0.02}^{+0.03}$ & $4.21$ \\
NGC~0628 & Sc-A & 9.0 & 6.5 & 10.30 & $-0.02$ & 208 & $7.14_{-0.07(-0.05)}^{+0.05(+0.05)}$ & $9.24_{-0.16(-0.06)}^{+0.09(+0.09)}$ & $-2.08_{-0.13(-0.16)}^{+0.13(+0.10)}$ & $0.80_{-0.10}^{+0.06}$ & $1.03_{-0.01}^{+0.01}$ & $5.39$ \\
NGC~5068 & Scd-AB & 9.0 & 26.9 & 10.10 & $-0.59$ & 30 & $7.17_{-0.03(-0.11)}^{+0.05(+0.07)}$ & $8.78_{-0.15(-0.08)}^{+0.15(+0.21)}$ & $-1.58_{-0.15(-0.31)}^{+0.12(+0.15)}$ & $0.60_{-0.08}^{+0.05}$ & $1.06_{-0.01}^{+0.01}$ & $7.10$ \\
NGC~2835 & Sc-B & 10.1 & 56.4 & 9.90 & $-0.40$ & 23 & $7.34_{-0.05(-0.14)}^{+0.02(+0.04)}$ & $8.87_{-0.18(-0.07)}^{+0.16(+0.28)}$ & $-1.58_{-0.12(-0.25)}^{+0.19(+0.27)}$ & $0.44_{-0.11}^{+0.08}$ & $1.08_{-0.01}^{+0.02}$ & $8.30$ \\
NGC~0598 & Scd-A & 0.9 & 58.0 & 9.65 & $-0.35$ & 19 & $7.52_{-0.03(-0.29)}^{+0.02(+0.06)}$ & $9.14_{-0.04(-0.00)}^{+0.13(+0.31)}$ & $-1.64_{-0.09(-0.45)}^{+0.07(+0.22)}$ & $0.85_{-0.05}^{+0.04}$ & $1.02_{-0.01}^{+0.01}$ & $8.95$ \\
\enddata
\tablecomments{Aliases for NGC~224, NGC~598, and NGC~5194 are M31, M33, and M51, respectively. The values of \tff, \tdep, and \eff\ are for SFR(FUV+22\micron), $H=100$~pc, $C<1$, and Galactic \aco. We provide the scatter of measurements ($+/-$ sign) as the range between 16th and 84th percentiles. The systematic uncertainties, defined as the largest difference between the median quantities from various assumptions, are written inside the parentheses. The standard errors of the median are very small ($\lesssim 0.01$ dex), and so not reported. Units of metallicity dependent \aco(1--0) are $M_\odot$ [K km s$^{-1}$ pc$^2$]$^{-1}$.}
\end{deluxetable*}

\acknowledgments

This Letter makes use of the following ALMA data: ADS/JAO.ALMA \#2013.1.00925.S \#2011.1.00650.S, and \#2013.1.00956.S. ALMA is a partnership of ESO (representing its member states), NSF (USA) and NINS (Japan), together with NRC (Canada), NSC and ASIAA (Taiwan), and KASI (Republic of Korea), in cooperation with the Republic of Chile. The Joint ALMA Observatory is operated by ESO, AUI/NRAO, and NAOJ. The National Radio Astronomy Observatory is a facility of the National Science Foundation operated under cooperative agreement by Associated Universities, Inc. 

The work of DU, AKL, and JS is partially supported by the National Science Foundation under Grants No. 1615105, 1615109, and 1653300. JMDK and MC gratefully acknowledge funding from the German Research Foundation (DFG) in the form of an Emmy Noether Research Group (grant number KR4801/1-1). JMDK gratefully acknowledges funding from the European Research Council (ERC) under the European Union's Horizon 2020 research and innovation programme via the ERC Starting Grant MUSTANG (grant agreement number 714907). ES acknowledges funding from the European Research Council (ERC) under the European Union`s Horizon 2020 research and innovation program (grant agreement No. 694343). GB is supported by CONICYT/FONDECYT, Programa de Iniciaci{\'o}n, Folio 11150220. FB acknowledges funding from the European Union`s Horizon 2020 research and innovation program (grant agreement No 726384 - EMPIRE). APSH is a fellow of the International Max Planck Research School for Astronomy and Cosmic Physics at the University of Heidelberg (IMPRS-HD). KK gratefully acknowledges support from grant KR 4598/1-2 from the DFG Priority Program 1573. The work of ECO is  supported by the NSF under grant No. 1713949. ER acknowledges the support of the Natural Sciences and Engineering Research Council of Canada (NSERC), funding reference number RGPIN-2017-03987. AU acknowledges support from Spanish MINECO grants ESP2015-68964 and AYA2016-79006. We acknowledge the usage of the Extragalactic Distance Database\footnote{\url{http://edd.ifa.hawaii.edu/index.html}} \citep{TULLY09}, the HyperLeda database\footnote{\url{http://leda.univ-lyon1.fr}} \citep{MAKAROV14}, the NASA/IPAC Extragalactic Database\footnote{\url{http://ned.ipac.caltech.edu}}, and the SAO/NASA Astrophysics Data System\footnote{\url{http://www.adsabs.harvard.edu}}.

\bibliographystyle{yahapj}

\end{document}